\documentstyle[aps,epsf,psfig]{revtex}
\newcommand{\be}[1]{\begin{equation}\label{#1}}
\newcommand{\ee}{\end{equation}}
\renewcommand{\vec}[1]{{\bf #1}}
\newcommand{\vecgr}[1]{\mbox{\boldmath{$ #1$}}}

\newcommand{\N}{\mbox{I}\!\mbox{N}}

\newcommand{\bino}[2]{\left(\begin{array}{c} #1\\#2\end{array}\right)}

\begin{document}
\title{
The autocorrelation function for spectral determinants of quantum graphs
}
\author{ Gregor Tanner}
\address{
School of Mathematical Sciences
\footnote{e-mail: gregor.tanner@nottingham.ac.uk}\\
University of Nottingham\\
University Park, Nottingham NG7 2RD, UK\\
and\\
Quantum Information Processing Group\\
Hewlett-Packard-Laboratories, Bristol\\
Filton Road, Stoke Gifford, Bristol BS34 8PQ, UK
}

\maketitle

\begin{abstract}
The paper considers the spectral determinant
of quantum graph families with chaotic classical
limit and no symmetries.
The secular coefficients of the spectral determinant 
are found to follow distributions
with zero mean and variance approaching a constant
in the limit of large network size.
This constant is in general different
from the random matrix result and depends on the classical limit.
A closed expression for this system dependent constant 
is given here explicitly in terms of the spectrum
of an underlying Markov process.\\
\noindent
{\normalsize submitted to {\em Journal of Physics A};\\
Version: 1st March 2002.}
\end{abstract}
% Version 13. March 2002

\section{Introduction}
\label{sec:sec1}
Quantum graphs and quantum networks have become a popular tool to
model quantum dynamics in mesoscopic systems in the diffusive regime
(Shapiro 1982, Chalker and Coddington 1988, see also
Dittrich 1996 and Janssen 1998 for recent review articles), as well as,
to study spectral statistics
(Kottos and Smilansky 1997, 1999, Pako\'nski et al 2001,
Ketzmerick et al 2000, Tanner 2001). In this paper, the autocorrelation
function of the spectral determinant
\be{specdet}
Z(\theta) = e^{-\frac{\rm i}{2}(N \theta + \varphi)}
\det\left({\bf 1} - e^{{\rm i}\theta} \vec{U}\right)=
e^{-\frac{\rm i}{2}(N \theta  + \varphi)} \sum_{n=0}^N a_n e^{{\rm i} n\theta}\ee
is studied after averaging over a suitable ensemble of quantum graphs
preserving the underlying 'classical' dynamics on the graph. Here,
$\vec{U}$ is a unitary $N \times N$ matrix which describes the quantum
evolution on the graph and the $a_n$'s denote the secular coefficients of the
characteristic polynomial. The phase factor in front of the determinant
ensures that $Z$ is a real function for real $\theta$ after setting
$\exp({\rm i}\varphi) = \det(-U)$
and using the basic property for the secular coefficients
\be{sec-coef} a_n = e^{{\rm i}\varphi} a^*_{N-n} \ee
which follows from the unitarity of $U$. The autocorrelation function of
(\ref{specdet}) has been studied by Ku\'s et al (1993)and Haake et al 
(1996) after averaging over circular-orthogonal,
-unitary and -symplectic ensembles of unitary matrices. Related results
for ensembles of Hermitian matrices have been given be Kettemann et al (1997).
All these approaches have in common that the average is taken over a
set of matrices too large to link the ensemble in the semiclassical limit
to a specific classical system. In this sense, the results are of random
matrix type. There is strong evidence, however, that the statistical
properties of spectra related to a fixed chaotic classical  system
already follow random matrix theory (RMT). This suggests that the size
of the ensembles considered can be reduced drastically without loosing
universality in the spectral statistics.

I will approach this problem by defining unitary matrix ensembles 
corresponding to a specific quantum graph. The autocorrelation
function of the spectral determinant is studied for these ensembles.
It will be shown
that the variance of the secular coefficients $a_n$ 
follow indeed the random matrix result for chaotic classical dynamics
up to a multiplicative factor. This factor is system dependent and
will be explicitly derived in section \ref{sec:sec4}. This improves the
formula given by Kettemann et al (1997) by carefully including
repetitions of periodic orbits leading to non-negligible
contributions. Related results have been reported by Keating and 
co-workers (Keating et al 1996) using periodic orbit approximations of the 
spectral determinant.

\section{Quantum graphs and Unitary Stochastic Ensembles}
\label{sec:sec2}
\begin{figure}
\centering
\centerline{
         \epsfxsize=5.5cm
         \epsfbox{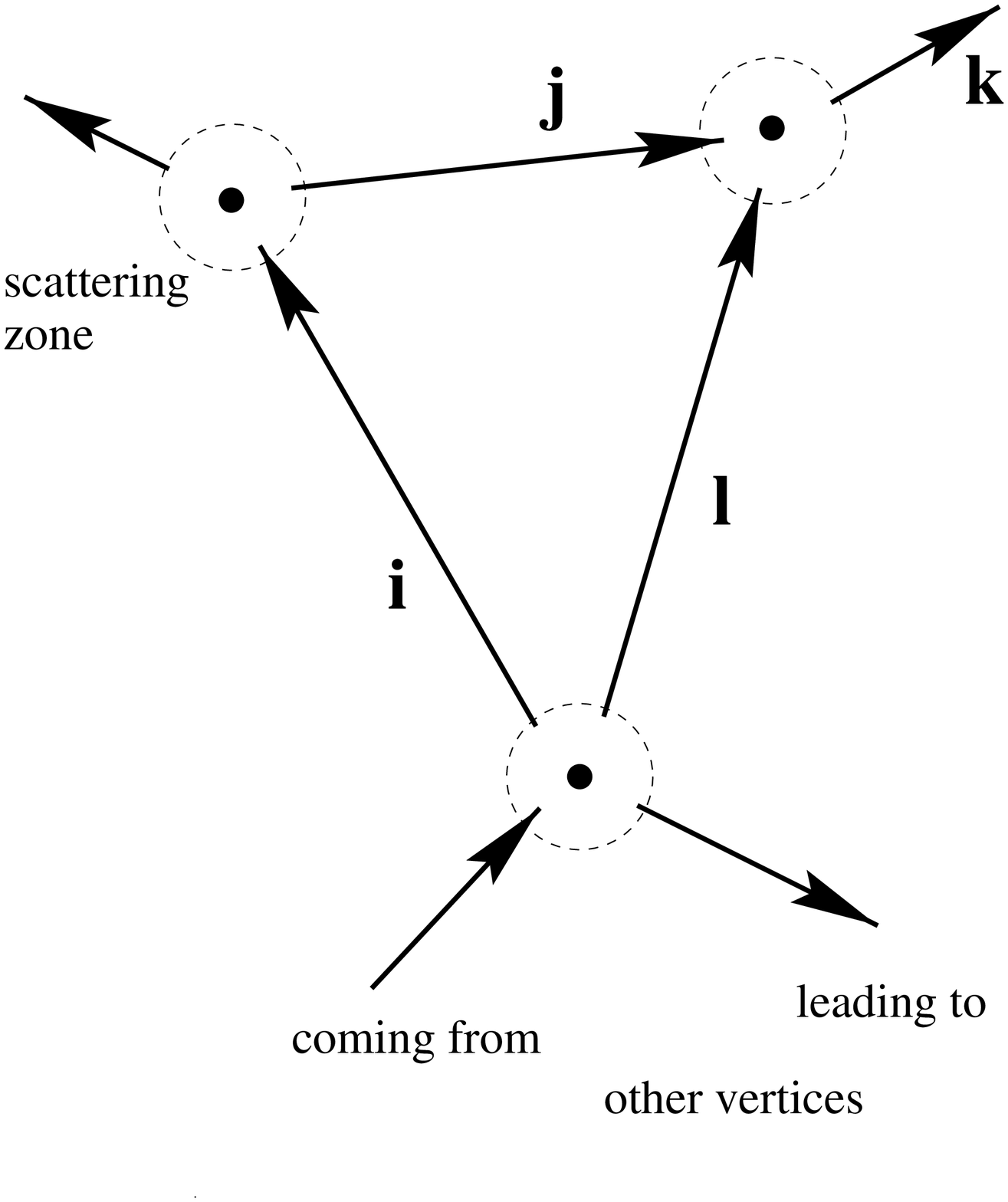}
         }
\caption[]{\small
Section of a typical directed graph; waves travel along directed edges
$i,j,k,l,\ldots$ undergoing scattering at the vertices.
}
\label{Fig:graph}
\end{figure}

In the following I will use a definition of quantum graphs which is slightly
more general than the one introduced by Kottos and Smilansky (1997). 
Consider a graph consisting of $N$ directed
edges or bonds connecting an unspecified
number of vertices. Free, one dimensional wave propagation occurs along edges
$i$ of length $L_i$.
Transitions from an edge $i$ to an edge $j$ is possible at a vertex if
and only if $i$ is an incoming edge and $j$ an outgoing edge at the
same vertex, see Fig.\ \ref{Fig:graph}. The scattering process at a given vertex is
described by a scattering amplitude $s_{ij}$, where $i$, $j$ label again
incoming and outgoing edges, respectively. To keep the set of possible
networks as general as possible, the local scattering amplitudes $s_{ij}$
will not be specified any further other than to fulfill basic properties
like probability conservation. (One may think of the scattering event taking
place at a given vertex to be due to interaction with a complex short-range
potential or a many-particle compound at that point).
%%%%%%%%%%%%%%%%%%%%%%%%%%%%%%%%%%%%%%%%%%%%%%%%%%%%%%%%%%%%%%%%%%%%%%%%%%%
%%%%%%%%%%%%%%%%%%%%%%%%%%%%%%%%%%%%%%%%%%%%%%%%%%%%%%%%%%%%%%%%%%%%%%%%%%%

The eigenstates (or resonances) of
such a network are given by stationary states of the system. These are obtained
by considering the $N$ - dimensional wave-vector $\vecgr{\Psi}_{\rm out}$
where the $i$-th component of $\Psi_{{\rm out}}$ represents the wave amplitude
on the $i$-th edge
when leaving the vertex. This vector is transformed into a set of
incoming waves $\vecgr{\Psi}_{\rm in}$ at the other end of the edge by
\[
\vecgr{\Psi}_{\rm in} = \vec{D}(k)\; \vecgr{\Psi}_{\rm out}
\qquad \mbox{with} \quad d_{ij} = e^{{\rm i} k L_i} \delta_{ij}
\]
where $k$ represents the wave number. The non-trivial part of the dynamics
is contained in the global scattering matrix $\vec{S}$ with
local scattering amplitudes $s_{ij}$ describing  transitions
between incoming and outgoing edges $i$ and $j$, that is,
\be{prop}
\vecgr{\Psi}_{\rm out} = \vec{S}\;\vecgr{\Psi}_{\rm in}
= \vec{S}\,\vec{D}(k)\; \vecgr{\Psi}_{\rm out} = \vec{U}(k) \; \vecgr{\Psi}_{\rm out}
\ee
which in turn defines the propagator
$\vec{U}(k)$ of the discrete wave dynamics on the network.
In the following, I will assume that the graph represents a closed system without
edges leading to infinity or absorbing vertices. This implies that
$\vec{S}$ and thus $\vec{U}$ are unitary. Stationary solutions are obtained for
wave numbers $k$ fulfilling the secular equation
\be{quant}
\det\left(\vec{1} - \vec{U}(k)\right) \stackrel{!}{=} 0 \, .
\ee
Certain restrictions on the topology of the graph and
its dynamical properties, that is, on the possible transition
probabilities $|s_{ij}|^2$, follow from the unitarity condition.
This point will be discussed later in more detail. I will furthermore
assume here that the scattering matrix $\vec{S}$ is independent of the
wavelength $k$.\\

Statistical properties of quantum systems are linked to the
complexity of an underlying classical dynamics. For quantum systems acting
on a continuous coordinate space with continuous or discrete time variable, the
classical dynamics revealed in the semiclassical limit is generated by
a set of Hamilton's equations of motion or for discrete times by a symplectic
map. For a quantum graph, the wave dynamics takes place on discretised
time and space variables and a connection to an underlying classical motion is
less obvious. The best one can achieve in these circumstances is to
identify the quantum propagation on a given graph with a Markov process. This
probabilistic 'classical' dynamics is generated by a
stochastic transition matrix $\vec{T}$ with matrix elements given as 
(Kottos and Smilansky 1997, Barra and Gaspard 2001, 2002)
\be{tran}  t_{ij} = |u_{ij}|^2 = |s_{ij}|^2 \ee
where $t_{ij}$ denotes the transition probability to jump from an edge $i$ to
an edge $j$ at a given vertex. Transition matrices which are derived from a
unitary matrix via condition
(\ref{tran}) are called unitary stochastic (Berman and Plemmons 1994).
A unitary stochastic matrix is also doubly stochastic,
that is, the rows and columns of $\vec{T}$ sum up to one. This implies, that
$\vec{T}$ has a largest eigenvalue equal to one with corresponding left
and right eigenvector $\frac{1}{N}(1,1,1,\ldots,1)$.
Note, that the converse is not true, that is, not every doubly stochastic
matrix is also unitary stochastic which leads to surprisingly complex
restrictions on the possible Markov processes which can be `quantised' in terms
of eqn.\ (\ref{tran}), see Pako\'nski and $\dot{\rm Z}$yczkowski (2002).

What constitutes the semiclassical limit of a quantum graph? The short wave
length limit $k \to \infty$ for fixed network size $N$ corresponds to a simple
scaling transformation, that is, the statistical properties of the spectrum are 
independent of $k$. Instead, one may construct a `semiclassical limit' for quantum graphs
by appropriately increasing the size of the graph. Such a limit should be
taken in a way which ensures convergence of the stochastic motion towards the
dynamics of a deterministic map within a given phase space resolution. This
semiclassical limit constitutes therefore at the same
time a classical (deterministic) limit of the stochastic dynamics. Sufficient
conditions for constructing such a limit will be formulated at a later stage.

The wavenumber $k$ defines a one-parameter family of unitary matrices $\vec{U}(k)$ for
a fixed quantum graph. A stationary solution exists for $k$-values, for which condition
(\ref{quant}) holds, that is,
whenever an eigenvalue $\exp({\rm i} \phi_i(k))$ of $\vec{U}(k)$ crosses the real line at $+1$.
Correlations in the $k$-spectrum are thus related to correlations in the
eigenphases spectrum $\phi_i(k)$ of $\vec{U}(k)$ for fixed $k$. Instead of studying
the $k$ spectrum directly, one can therefore look at the 
statistical properties of the eigenphases of $\vec{U}(k)$
for fixed $k$ and then average over $k$.

In the generic case, when the set of lengths $L_i$ are all rationally
independent, the one-parameter family $\vec{U}(k)$ covers uniformly a subspace of the
unitary group ${\cal U}(N)$ with the topology of an (at most) $N$ dimensional torus.
That is, the $k$ average can be replaced by an ensemble average over the ensemble
${\cal U}_T$ of unitary matrices defined as
\[ {\cal U}_T =
\{ \vec{U} \in {\cal U}(N)| \, \vec{U} = \vec{S}\, \vec{D} \;\;
\mbox{with}\; d_{ij} = e^{{\rm i} \varphi_i} \delta_{ij}, \;\; \varphi_i \in [0,2\pi)\;
\forall i=1,\ldots N\}
\]
where the scattering matrix $\vec{S}$ is kept fixed and the integration measure for the
ensemble is given by
\[
d\mu = \prod_{i=1}^N \frac{d\varphi_i}{2\pi} \, .
\]
Note that the ensemble average corresponds to averaging over the spectrum of {\em a specific
quantum system with fixed classical dynamics}. All the unitary matrices forming
the ensemble are indeed linked to the same stochastic
process on the graph described by a unitary stochastic transition matrix $\vec{T}$.
I will therefore denote these ensembles as {\em Unitary Stochastic Ensembles
${\cal U}_T$} (USE) (Tanner 2001).
\footnote{It should be noted that a USE is in general not uniquely defined by a unitary stochastic
matrix $\vec{T}$ for $N > 2$; there are in general several unitary matrices
$\vec{S}$ not connected by multiplication with diagonal matrices
which corresponds to the same transition matrix $\vec{T}$. The statistical properties of
all these ensembles are, however, expected to be identical in the limit $N\to\infty$.}

It has been argued in Tanner (2001) that correlations in the eigenphases spectrum
after ensemble averaging depend crucially on the spectral gap
of $\vec{T}$, that is, on the distance of the second largest eigenvalue of $\vec{T}$
from the unit circle. More precisely, it was conjectured that the spectral correlations
follow random matrix theory in the classical limit if the gap decreases slower than
$1/N$ for $N \to \infty$.
I will in the following study a different statistical measure in more detail, namely the
auto-correlation function for spectral determinants.

\section{The autocorrelation function for the spectral determinant}
\label{sec:sec3}
The averaged autocorrelation function $C(x)$ of the spectral determinant
(\ref{specdet}) may be defined in terms of the
generating function for the square moduli of the secular coefficients, that is,
\be{gen-fun}
P(z) =
\sum_{n=0}^N <|a_n|^2>_{{\cal U}_T} z^n \, ,
\ee
where the average is taken over a unitary stochastic ensemble as defined above. One obtains
\be{corr}
\tilde{C}(\omega) = e^{-{\rm i}\, \frac{N}{2}\omega} \frac{
<\int_0^{2\pi} {\rm d}\theta\; Z(\theta+\frac{\omega}{2})\,
Z^*(\theta-\frac{\omega}{2})>_{{\cal U}_T} }
{<\int_0^{2\pi} {\rm d}\theta\, |Z(\theta)|^2 >_{{\cal U}_T} }
 = e^{-{\rm i}\, \frac{N}{2}\omega} \frac{P(e^{{\rm i} \omega})}{P(1)} \; .
\ee
It is in general convenient to rescale the argument such that the correlation
function is periodic with period one, that is, one considers
$C(x) = \tilde{C}(\frac{2 \pi x}{N})$. The symmetry property
(\ref{sec-coef}) ensures
\be{symm} <|a_n|^2> = <|a_{N-n}|^2> \quad \mbox{ and thus} \quad C(x) = C(1-x)\, . \ee

Averages will always be taken over USE's unless stated otherwise and I will
drop the suffix $<.>_{{\cal U}_T}$ from now on. One obtains
immediately
\[ <a_n> = 0 \quad \forall n=0,\ldots, N \]
when averaging over the phases $\varphi_i$. The variance of the secular
coefficients $<|a_n|^2>$ has been calculated in the context of RMT
by Haake et al (1996). One obtains in particular after averaging over the whole
unitary group ${\cal U}(N)$ (CUE), over the ensemble of symmetric unitary
matrices (COE) or over the ensemble of diagonal unitary matrices (Poisson)
\be{RMT}
<|a_n|^2> = \left\{\begin{array}{ll} 1 & \mbox{CUE}\\
                             1 + \frac{n(N-n)}{N+1}\; & \mbox{COE}\\
                             \bino{N}{n}\; & \mbox{Poisson}
                  \end{array} \right. \, .
\ee
Note, that the CUE and COE
distributions as well as the CSE distribution not shown here (Haake et al 1996)
converge to a limiting distribution for large $N$ with respect to the parameter
$\tau = n/N$ whereas no such asymptotic limit exists in the Poisson case.
It will be shown in the following that the ensemble average for USE's without
symmetries may deviate from the CUE result above. \\

Using Laplace expansion of the determinant, one writes
\[ \det\left({\bf 1} - z \vec{U}\right)=
1 + \sum_{n=1}^{N} z^n \sum_{\alpha\in \Gamma(n,N)} (-1)^n \det(\vec{U}[\alpha])
\]
where $\Gamma(n,N)$ denotes the set of integer vectors with
\[
\Gamma(n,N) =
\{ \alpha\in\N^n | 1\le \alpha_1< \alpha_2 <\ldots<\alpha_n\le N\}
\]
and $\vec{U}[\alpha]$ is the $n \times n$ matrix obtained from $\vec{U}$
by taking only the rows and columns of $\vec{U}$ with indices
$\alpha_1, \ldots, \alpha_n$, that is,
\[ u[\alpha]_{ij} = u_{\alpha_i\alpha_j}.\]
The variance of the distribution of secular coefficients can thus be written
as a double sum over all possible subgraphs spanned by the edges
$\alpha = (\alpha_1, \ldots, \alpha_n)$,
\begin{eqnarray} \nonumber
<|a_n|^2> &=&
\sum_{\alpha, \alpha' \in \Gamma(n,N)}
<\det(\vec{U}[\alpha]) \det(\vec{U}^{\dagger}[\alpha'])> \\
&=& \sum_{\alpha \in \Gamma(n,N)}
<\det(\vec{U}[\alpha]) \det(\vec{U}^{\dagger}[\alpha])> \, .
\label{diag1}
\end{eqnarray}
In the last step, the fact that contributions
with $\alpha \ne \alpha'$ vanish identically when taking the ensemble
average has been exploited.
Using Laplace expansion again one writes each of the
sub-determinants  in (\ref{diag1}) as a sum over permutations $\pi$ of the
row (or column) indices, that is,
\[
\det(\vec{U}[\alpha]) = \sum_{\pi} (-1)^{\pi} u[\alpha]_{\pi} =
\sum_{\pi} (-1)^{\pi} \prod_{i=1}^{n} u[\alpha]_{i,\pi(i)} \, .
\]
The sum over permutations may be interpreted as a sum over closed
(periodic) paths on the subgraph $\alpha$ which visit each edge
$\alpha_i$, $i=1,\ldots n$ exactly once. These periodic paths will
be called covering orbits in what follows; they are not
necessarily connected. The product of determinants in (\ref{diag1})
now takes on the form
\[
<|a_n|^2> = \sum_{\alpha \in \Gamma(n,N)}
<\sum_{\pi, \pi'} (-1)^{\pi+\pi'} u[\alpha]_{\pi} u[\alpha]^*_{\pi'}> \, .
\]
Periodic orbit pairs with $\pi = \pi'$  will obviously survive the
ensemble average and may form an important contribution to the double sum.
Splitting these diagonal contributions from the off-diagonal contributions
and assuming that there are no systematic periodic orbit degeneracies due
to symmetries, one obtains
\begin{eqnarray} \nonumber
\frac{1}{n!} \frac{d^n}{dz^n} P(z) \left|_{z=0}\right. =
<|a_n|^2>
&=& \sum_{\alpha \in \Gamma(n,N)}
\sum_{\pi} t[\alpha]_{\pi} + <\sum_{\pi \ne \pi'}
(-1)^{\pi+\pi'} u[\alpha]_{\pi} u[\alpha]^*_{\pi'}> \\
&=& \sum_{\alpha \in \Gamma(n,N)}
\mbox{per} (\vec{T}[\alpha]) + <\sum_{\pi \ne \pi'}
(-1)^{\pi+\pi'} u[\alpha]_{\pi} u[\alpha]^*_{\pi'}> \, . \nonumber
\end{eqnarray}
Here, $t[\alpha]_{\pi} = |u[\alpha]_{\pi}|^2$ denotes the product
of transition probabilities along the periodic path. Furthermore,
$\vec{T}[\alpha]$ is the sub-matrix obtained from the
rows and columns $\alpha_1, \ldots \alpha_n$ of
the stochastic transition matrix $\vec{T}$ and
$\mbox{per}(\vec{T}[\alpha])$ is the permanent of this matrix. Note,
that the permanent of a square matrix $\vec{A}$ is defined as
\[
\mbox{per}(\vec{A}) = \sum_{\pi} \prod_{i=1}^{n} a_{i,\pi(i)}\, .
\]
A detailed account of the properties of permanents may be found in
Minc (1978), Berman and Plemmons (1994) or Brualdi and Ryser (1991).

One finally obtains for the generating function (\ref{gen-fun})
\be{diag2}
P(z) =
\sum_{n=0}^N <|a_n|^2> z^n =
\mbox{per}\left(\vec{1} + z \vec{T}\right) + \mbox{non-diagonal contributions}\, .
\ee
The autocorrelation function of the spectral determinant can thus
be linked to the permanent of sub-matrices
of the stochastic transition matrix $\vec{T}$. This is in
analogy to the connection between the spectral correlation function
and the traces of powers of $\vec{T}$ (Tanner 2001).
Note, that additional contributions to the diagonal approximation have
to be taken into account in the presence of symmetries as for example
time reversal symmetry; I will not consider this case here
any further.

Two questions arise immediately: firstly, what is the range of validity
of the diagonal approximation, and secondly, how can one
extract at least asymptotic results in the large $N$ limit from
the expression (\ref{diag2}) by for example
making a connection to the spectrum of the transition matrix $\vec{T}$?
Especially the last point turns out to be quite tricky; dealing with
permanents is a very hard problem in general (unlike calculating
determinants), the effort of computing a permanent increases
indeed faster than exponential with the matrix size.
\footnote{Calculating the permanent of a general matrix is an
NP complete problem in the language of algorithmic complexity, that
is, there is no (known) polynomial algorithm to compute the
permanent; see Brualdi and Ryser (1991), as well as Minc (1978)
for a description of the best known algorithms.}
I will give answers to both the points raised above in the next section.

\section{Asymptotic results for chaotic dynamics on graphs}
\label{sec:sec4}
When studying the autocorrelation function in the limit of large
network size, one first has to specify, how to take this limit, that
is, how to choose families of graphs which may be said to have a well
defined classical and thus semiclassical limit for $N\to \infty$.
Taking a very conservative point of view, I will demand that an
increase of the size of the network has to be done in a way which leaves
the probabilistic dynamics of the Markov process and thus the
periodic orbit structure invariant up to a certain cutoff time
$n_c(N)$; the cutoff time itself needs to increases with
$N$, one can typically achieve $n_c \sim \log N$. Such
a definition ensures automatically that typical measures of chaos
like the topological entropy $h_t$, the K - entropy or the spectral
gap converge for large $N$. Only graph
families without symmetries and with a finite gap and such exponential
decay of correlations will be considered in what follows.

The transition matrices of such families of graphs with chaotic
classical limit become increasingly sparse for large $N$; the
number of non-zero matrix elements increases typically not faster than
$N$. This in turn implies that most of the subgraphs $\alpha$ of size
$n$ entering the sum (\ref{diag1}) contain no or at least one
covering orbit for small $n$. (I will make no
distinction between connected and non-connected orbits here.) The
diagonal approximation is obviously exact in these cases.
Non-diagonal contributions will become important for $n$ values for which
two or more covering orbits exist in a typical subgraph. A
transition between these two regimes occurs roughly at times which
coincide with the radial size of the network, that is, with the
mean time to reach each edge from every other edge. This transition time
$n_t$ is of the order
\[ n_t(N) \sim \frac{\log N}{h_t}\]
where $h_t$ is the topological entropy. The diagonal approximation can thus
be estimated to be valid for coefficients $<|a_n|^2>$ with
$n < n_t$.

A connection between the permanent appearing in (\ref{diag2}) and the
spectrum of the transition matrix can be established using similar arguments.
Consider first the identity
\be{zeta}
\det(\vec{1} - z \vec{T})  = \prod_{p} (1 - z^{n_p} t_{p})
\ee
for determinants,
where the product runs over {\em all} primitive periodic orbits
of the graph, that is, connected periodic paths not including
repetitions. Here, $n_p$ denotes the length of the orbit and the weight
$t_p$ corresponds to the product of transition probabilities along the
orbit. One might now be tempted to just replace the minus sign
on the right hand side in (\ref{zeta}) by a plus sign, that is, to
consider the function
\be{zeta+}
\zeta_+(z) = \prod_{p} (1 + z^{n_p} t_{p}) = \sum_{n=0}^{\infty} z^n c_n
\ee
to obtain a periodic orbit expression for
$\mbox{per}\left(\vec{1} + z \vec{T}\right)$.
The product in (\ref{zeta}) does, however, include self-intersecting
periodic paths on the graph, that is, periodic orbits which visit the
same edge more than once. It is one of the magic properties of the
determinant, that, after expanding the product in (\ref{zeta}), contributions
of intersecting orbits cancel exactly, leaving only terms containing
orbits and products of orbits, which have no edge in common. The
same is obviously not true for the expression (\ref{zeta+}).
One can argue, however, that short periodic orbits on the graph
are in general irreducible, ie, they do not self-intersect
(Bogomolny 1992, Tanner and Wintgen 1992). Furthermore, taking the classical limit
in the sense described above corresponds to enhancing the phase space
resolution, that is, self-intersecting orbits turn eventually into
irreducible orbits when increasing $N$. Again, self-intersections are likely
to occur only for times $n$ larger than the transition time $n_t$. The
coefficients of a polynomial expansion of the permanent in
(\ref{diag2}) can thus be approximated by the coefficients of $\zeta_+(z)$
exactly in the regime, where the diagonal approximation holds, that is,
\be{equiv}
<|a_n|^2> \approx \frac{1}{n!} \frac{d^n}{dz^n} \mbox{per}\left(\vec{1} + z \vec{T}\right)
\left|_{z=0} \right.
\approx \frac{1}{n!} \frac{d^n}{dz^n} \prod_{p} (1 + z^{n_p} t_{p})
\left|_{z=0} \right.
\quad \mbox{for} \; n < n_t \sim \frac{\log N}{h_t} \, .
\ee
The zeta function (\ref{zeta+}) can be linked to the transition matrix
by the following calculation, ie,
\begin{eqnarray}\nonumber
\zeta_+ &=& \prod_{p} (1 + z^{n_p} t_{p})
        = \exp\left[\sum_{p} - \sum_{r=1}^{\infty}
    \frac{(-1)^r}{r} z^{r n_p} t^r_{p}\right] \\ \nonumber
        &=& \exp\left[\sum_{r=1}^{\infty}
    \frac{1}{r} \left(\sum_p (z^{n_p} t_{p})^r
    - \sum_p (z^{n_p} t_{p})^{2 r}\right)
    \right] \\
    &=& \frac{\det\left(\vec{1} - z^2 \vec{T}^{(1)}\right)}
    {\det\left(\vec{1} - z \vec{T}\right)} \, .
    \label{zet-det}
\end{eqnarray}
The matrix $\vec{T}^{(1)}$ is defined by squaring the matrix-elements
of $\vec{T}$, that is, $t^{(1)}_{ij} = t^2_{ij}$.
Equation (\ref{zet-det}) is the main result of this paper. It differs from the
relation given by Kettemann et al (1997) by having a determinant also in the
numerator; this term arises by including repetitions of periodic orbits in
a consistent manner and must not be neglected. It
leads to significant contributions in the expansion of $\zeta_+$ and
thus to the coefficients $<|a_n|^2>$ in general. These additional contributions 
have also been noted by Keating and co-workers (1996). Their results, derived
from semiclassical periodic orbit expansions of the spectral determinant, are
more general in the sense that they apply for generic quantum systems. The resulting
periodic orbit formula equivalent to (\ref{zet-det}) could, however, not be linked to
classical operators. The periodic orbit formulas are notoriously difficult to
calculate and specific values for the secular coefficients could indeed not be
obtained so far.

Here, one may proceed as follows;
after using Newton's method (Kettemann et al 1997) to expand the two determinants in
(\ref{zet-det}) in terms of traces, one recovers the coefficients $c_n$ of
$\zeta_+$ in eq.\ (\ref{zeta+}) recursively by applying standard formulas for
fractions of power series (Gradshteyn and Ryzhik 1994). The $c_n$'s converge
exponentially fast to a
finite value for large $n$ for $\vec{T}$ matrices with a finite spectral gap.
The asymptotic behaviour is in particular determined by the residium of
$\zeta_+(z)$ at the leading singularity at $z=1$. It follows from the
relation (\ref{equiv}) which is valid over a range of $n$ values increasing with $N$
that the coefficients $<|a_{\tau}|^2>$ converge to a constant value $\alpha$ for fixed
$\tau = n/N$ and $N\to \infty$ given as
\be{asym}
\alpha = \lim_{N\to\infty} <|a_{\tau}|^2> = \lim_{N\to\infty} \mbox{Res}_{z=1}
\frac{\det\left(\vec{1} - z^2 \vec{T}^{(1)}\right)}
{\det\left(\vec{1} - z \vec{T}\right)}\; .
\ee
This result differs from the CUE result (\ref{RMT}) insofar that the constant
$\alpha$ is in general not
equal to one. It depends furthermore on the asymptotic limit of the eigenvalue spectra of both
$\vec{T}$ and $\vec{T}^{(1)}$ and thus on the classical limit of the family of graphs considered.
Let me demonstrate this point with the help of a simple example.\\
\noindent
{\em Example.} Consider the sawtooth map on the unit interval with $k$ legs of slope $k$. The map
has a simple Markov partition whose $l$-th refinement divides the unit interval into $k^l$
subintervals of equal length. The Markov partition defines a transition matrix $\vec{T}_k(N)$ of
dimension $N=k^l$ with $k^{l+1}$ nonzero matrix elements all equal to $1/k$. The so defined transition
matrices are unitary stochastic corresponding to a graph where each vertex has $k$ incoming and
outgoing edges. The spectrum of $\vec{T}_k$ and $\vec{T}^{(1)}_k$ consist of a single non-zero
eigenvalue being 1 and $1/k$, respectively, independent of the order of refinement $l$. Transition
matrices for fixed $k$ form a family with well defined classical limit. Following
eqs.\ (\ref{zet-det}), (\ref{asym}) one expects
for the coefficients of the autocorrelation function
\[ <|a_0|^2> = 1,\; <|a_1|^2> = 1,\; \alpha= <|a_n|^2> = 1-\frac{1}{k}
\quad \mbox{for} \quad 1<n<\frac{N}{2}\]
in the limit $N\to \infty$. (The function $\zeta_+$ does of course not obey the
symmetry conditions (\ref{zet-det}).)

In the next section, I will consider a non-trivial example in more detail, namely
so-called binary graphs (Tanner 2000, 2001).

\section{Numerical results}
\label{sec:sec5}
The model systems considered are graphs with an even number of edges and transition
matrices of the form
\be{adj_bin} t_{ij} = \left\{ \begin{array}{ll}
     \frac{1}{2}(\delta_{2i,j} + \delta_{2i+1,j}) &\mbox{for} \;\; 0 \le i < \frac{N}{2}\\
     \frac{1}{2}(\delta_{2i-N,j} + \delta_{2i+1-N,j}) &\mbox{for} \; \; \frac{N}{2} \le i < N
                    \end{array} \right . \quad i = 0, \ldots, N-1\; .
\ee
The graphs represent for dimensions $N = 2^l$ the Markov partition of a two-leg
sawtooth map with complete binary symbolic dynamics. This is a special
case of the example above for $k = 2$. I will refer to graphs of
the form (\ref{adj_bin}) as binary graphs. The transition matrices
(\ref{adj_bin}) are unitary stochastic; one can furthermore show, that
binary graphs with dimensions $N = p\, 2^l$ for fixed $p$ being an odd integer
form a family with a well defined classical limit. The spectral gap is
$\Delta = 1/2$ for $p \ne 1$ and the spectrum of $\vec{T}$ (as well as
$\vec{T}^{(1)}$) depends on $p$, but not on the order $l$.

\begin{figure}
\centering
\centerline{
         \epsfxsize=11cm
         \epsfbox{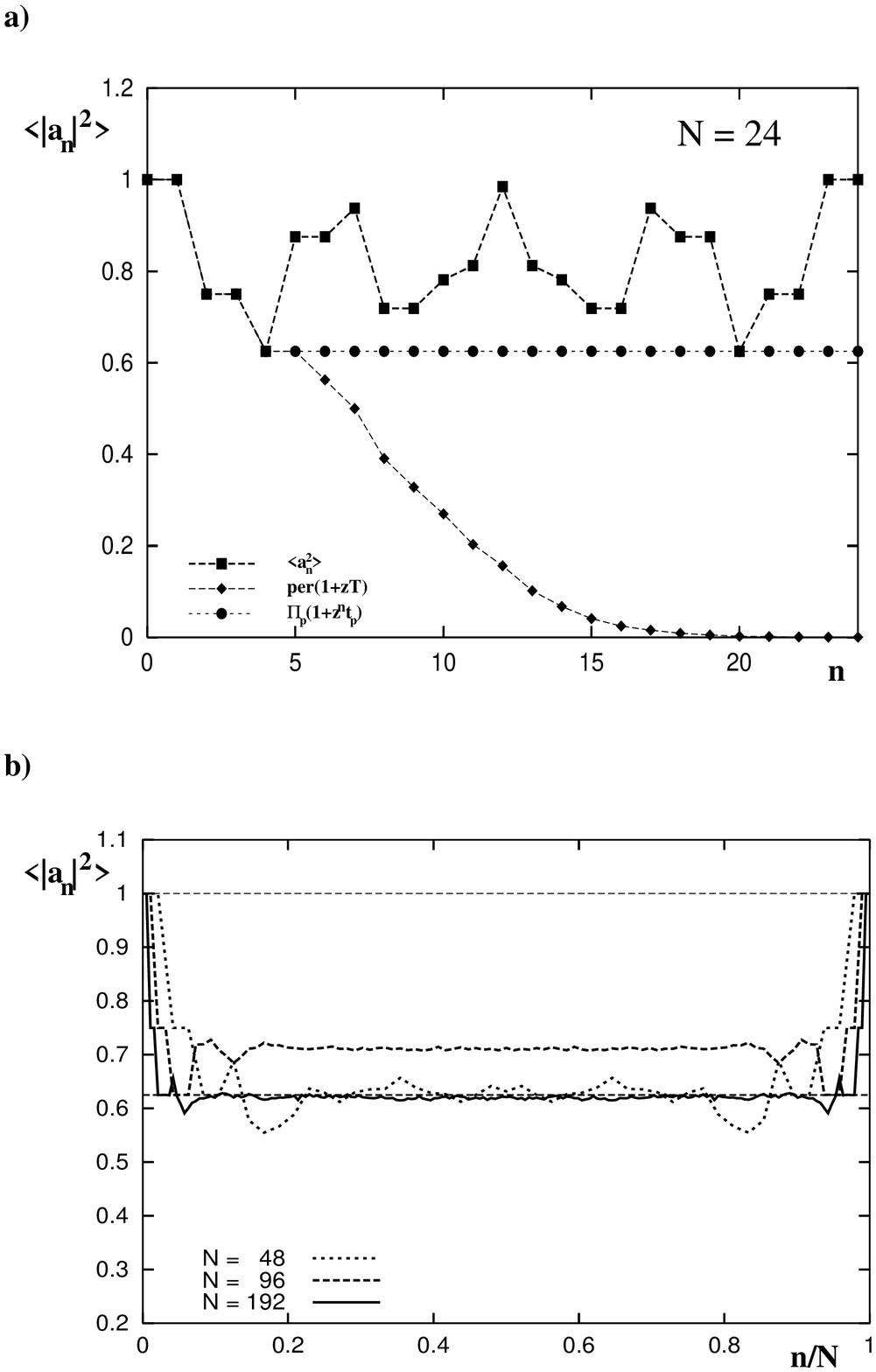}
         }
\caption[]{\small
Coefficients of the autocorrelation function for the binary graph family with
$p=3$;\\
a) coefficients $<|a_n|^2>$ compared with the diagonal approximation
obtained from the coefficients of $\mbox{per}\left(\vec{1} + z \vec{T}\right)$
and the periodic orbit formula $\zeta_+ = \prod_{p} (1 + z^{n_p} t_{p})$
for $N = 24$;
the asymptotic result for this family is 0.625;\\
b) convergence to the asymptotic result for members of the same family with
larger network size; the random matrix result is $<|a_n|^2> = 1$.
}
\label{Fig:cor3}
\end{figure}

In Fig.\ \ref{Fig:cor3}a, the secular coefficients $<|a_n|^2>$ obtained from
(\ref{diag1}) are shown together with the coefficients
of per$(\vec{1}+z \vec{T}$) and $\zeta_+(z)$ for a member of the $p = 3$ family
with dimension $N = 24$. The coefficients of the various approximations all
coincide for $n \le 5 \sim n_t = \log 24/\log 2$.  The coefficients of the
permanent function decrease rapidly thereafter, whereas those of the
periodic orbit product approach a constant (which is 0.625 for this particular
family).  The coefficients of the autocorrelation function still fluctuate
wildly for such a small $N$ value. Fig.\ \ref{Fig:cor3}b shows
various members of the same family for
larger $N$ values. The fluctuations decrease considerably and convergence to
the limiting value $\alpha= 0.625$ is observed.
The asymptotic value depends on the family and thus on $p$; one observes in
particular convergence to the asymptotic value $\alpha= 1/2$ for $p = 1$ as
predicted in the example in the previous section.\\

\noindent
{\large \bf Acknowledgments}\\
\noindent

I would like to thank Prot Pako\'nski and Karol $\dot{\rm Z}$yczkowski
for stimulating discussions. I am also grateful for the hospitality
experienced during numerous stays at the Hewlett--Packard Laboratories
in Bristol and for financial support from the Nuffield Foundation and the
Royal Society.

%%%%%%%%%%%%%%%%%%%%%%%%%%%%%%%%%%%%%%%%%%%%%%%%%%%%%%%%%%%%%%%%%%%%%%%%%%%
%%%%%%%%%%%%%%%%%%%%%%%%%%%%%%%%%%%%%%%%%%%%%%%%%%%%%%%%%%%%%%%%%%%%%%%%%%%

%
%
%
%

\end{document}